\documentclass[12pt]{article}
\usepackage[pdftex]{graphicx}
\usepackage{amssymb}
\usepackage{amsmath}

\newcommand{\rf}[1]{(\ref{#1})}
\newcommand{\beq}{\begin{equation}}
\newcommand{\eeq}{\end{equation}}
\newcommand{\bea}{\begin{eqnarray}}
\newcommand{\eea}{\end{eqnarray}}

\newcommand{\e}{\mbox{e}}

\newcommand{\g}{\gamma}

\newcommand{\lam}{\lambda}
\newcommand{\Lam}{\Lambda}
\newcommand{\ep}{\varepsilon}

\newcommand{\del}{\delta}

\newcommand{\ra}{\rangle}
\newcommand{\la}{\langle}
\newcommand{\prt}{\partial}

\newcommand{\cD}{{\cal D}}

\newcommand{\cH}{{\cal H}}

\newcommand{\cL}{{\cal L}}

\newcommand{\tL}{{\tilde{\Lam}}}

\newcommand{\hH}{{\hat{H}}}

\newcommand{\hL}{{\hat{L}}}

\newcommand{\pd}[2]{\frac{\partial #1}{\partial #2}}

\newcommand{\pg}{\pi^{\g}}
\begin{document}

\begin{center}
\vspace{24pt}
{ \large \bf 2d CDT is 2d Ho\v{r}ava-Lifshitz quantum gravity}

\vspace{30pt}

{\sl Jan Ambj\o rn}$\,^{a,d}$,
{\sl Lisa Glaser}$\,^{a}$,
{\sl Yuki Sato}$\,^{b}$
and {\sl Yoshiyuki Watabiki}$\,^{c}$

\vspace{48pt}
{\footnotesize

$^a$~The Niels Bohr Institute, Copenhagen University\\
Blegdamsvej 17, DK-2100 Copenhagen \O , Denmark.\\
email: ambjorn@nbi.dk, glaser@nbi.dk\\

\vspace{10pt}

$^b$~Department of Physics, Nagoya University, \\
Nagoya 464-8602, Japan.\\
email: ysato@th.phys.nagoya-u.ac.jp

\vspace{10pt}

$^c$~Tokyo Institute of Technology,\\ 
Dept. of Physics, High Energy Theory Group,\\ 
2-12-1 Oh-okayama, Meguro-ku, Tokyo 152-8551, Japan\\
{email: watabiki@th.phys.titech.ac.jp}

\vspace{10pt}

$^d$~Institute for Mathematics, Astrophysics and Particle Physics\\ 
Radbaud University Nijmegen, \\Heyendaalseweg 135,
6525 AJ, Nijmegen, The Netherlands 
}

\vspace{48pt}
\end{center}

%\addtolength{\baselineskip}{0.20\baselineskip}
%\vspace{2cm}

\begin{center}
{\bf Abstract}
\end{center}

Causal Dynamical Triangulations (CDT) is a  
lattice theory  where aspects of 
quantum gravity can be studied. Two-dimensional CDT can 
be solved analytically and the continuum (quantum) Hamiltonian obtained.
In this article we show that this continuum Hamiltonian is
the one obtained by quantizing two-dimensional projectable 
Ho\v rava-Lifshitz gravity.

\noindent 
\vspace{12pt}
\noindent

\vspace{24pt}
%\noindent
%PACS:  \\
%Keywords: 

\newpage

%%%%%%%%%%%%%%%%%%%%%%%
%%%%%%%%%%%%%%%%%%%%%%%
\section{Introduction}\label{intro}
%%%%%%%%%%%%%%%%%%%%%%%
%%%%%%%%%%%%%%%%%%%%%%%

Causal Dynamical Triangulations (CDT) provides a non-perturbative 
lattice regularization for a quantum theory of  geometries which 
are characterized by allowing a time foliation (for a recent review 
see \cite{Ambjorn:2012jv}). While originally formulated for spacetimes
with Lorentzian signature, the formulation is such that 
one can rotate to Euclidean signature and perform the 
path integral there.  In this sector one can prove that 
the lattice theory has a transfer matrix which is 
reflection positive. One thus expects that a continuum 
theory obtained from the lattice theory will have a unitary 
time-evolution. 

The features just mentioned are by large the same features 
as Peter Ho\v rava imposed on what is now known as Ho\v rava-Lifshitz
gravity \cite{Horava:2009uw}. 
Thus CDT should provide a lattice regularization
of Ho\v rava-Lifshitz gravity (HL). This was first suggested in \cite{spectral1}
where it was noted that the spectral dimension in HL gravity
agreed with the one measured in CDT \cite{spectral2}. 
Indeed, there is good evidence that 
CDT can provide  a regularization for HL gravity \cite{evidence}.
So far this evidence is based 
on numerical simulations in three and four dimensions\footnote{
Of course there is also in addition the possibility 
that one obtains a quantum theory of standard general relativity (GR) 
for special values of the bare lattice coupling constants, and the 
search for such ``isotropic'' points is an important part of 
the CDT program.}. Unfortunately
it  is difficult to solve the lattice theory analytically 
and  equally difficult
to address details of the continuum renormalizable quantum theories in these 
dimensions. However, the two-dimensional theory provides a simple
test: the lattice theory can be solved analytically (\cite{al}) 
and in 2d we can analyze the corresponding continuum HL quantum theory and show
that it agrees with 2d CDT.

\section{Projectable Ho\v{r}ava-Lifshitz gravity}\label{PHL} 

In HL it is assumed that spacetime has a time foliation and 
that the theory is invariant under spatial diffeomorphisms 
as well as time redefinitions, i.e.\ so-called  
foliation preserving diffeomorphisms:
\begin{equation}\label{j1}
t \to t + \xi^{0}(t), \ \ \ x^i \to x^{i} + \xi^i (t,x),
\end{equation}
where $i=1,\ldots,d$,  $d$ being the dimension 
of space.  We will be interested in the very simplest version of 
this theory which contains at most second order derivatives of 
the metric. It is convenient to use the ADM-decomposition of the 
metric 
\begin{equation}\label{j2}
g_{\mu \nu}(x,t) dx^{\mu} dx^{\nu} = -N(x,t)^{2} dt^2 + 
g_{ij}(x,t) (dx^i+ N^{i}(x,t) dt)(dx^j+ N^{j}(x,t) d t),
\end{equation} 
where $N(x,t)$ is the lapse function and $N^{i}(x,t)$ is the shift vector.
The  action is   given by
\begin{equation}\label{j3}
I = \frac{1}{\kappa} \int  dt \; 
d^{d}x \sqrt{g} N \Big[( K_{ij}K^{ij} - \lambda K^2) +R-2 \Lambda  \Big]. 
\end{equation}
Here $g_{ij}$ is the spatial metric on the foliation, $R$ the 
corresponding scalar curvature, and   
$K_{ij}$ is the extrinsic curvature defined by the lapse 
function $N$ and the shift vector $N^i$ as
\begin{equation}\label{j4}
K_{ij} = \frac{1}{2N}\, 
\Big( \partial_0 {g}_{ij} - \nabla_{i}N_{j} - \nabla_{j}N_{i} \Big),
\end{equation}
where $\nabla_{i}$ is the covariant derivative associated with $g_{ij}$,
and $K = g^{ij}K_{ij}$. 
Finally $\kappa$ is the gravitational 
coupling constant analogous to the Newton constant in 
GR, $\Lambda$ the cosmological constant  
and  $\lambda$ a dimensionless parameter. If $\lambda =1$ 
one recovers GR and in  this case 
$I$ is just the integrated intrinsic scalar curvature, i.e.\
the standard GR action. However, for $\lambda \neq 1$ 
$I$ is only invariant under the foliation preserving 
coordinate transformations (\ref{j1}).

As is standard in the ADM decomposition
one can (up to boundary terms) rewrite the action as 
\begin{equation}\label{j5}
  I= \int dt \int d^dx\, \Big[ \pi^{ij} \partial_0 g_{ij} 
-N {\cal{H}} -N_i {\cal H}^i\Big],
\end{equation}  
where 
\begin{equation}\label{j5a}
{\cal H} = \frac{\kappa}{\sqrt{g}}\, 
\Big(\pi^{ij}\pi_{ij} -\frac{\lambda}{d\lambda-1}\pi^2\Big)-
\frac{\sqrt{g}}{\kappa} (R-2\Lambda),
~~~~~{\cal H}^i = -2 \nabla_j\pi^{ij}, 
\end{equation}
and where the canonical momenta are 
\begin{equation}\label{j6}
\pi^{ij} = \frac{\prt \cL}{\prt (\prt_0 g_{ij})} = 
\frac{\sqrt{g}}{\kappa} \,(K^{ij} -\lam g^{ij}K),~~~~~\pi=g_{ij}\pi^{ij}.
\end{equation}
Here $\cL$ is the Lagrangian density corresponding to the action $I$.
$\pi^{ij}$ is a tensor density and  $\nabla_j\pi^{ij}$ is the standard 
covariant derivative of a tensor density.
${\cal H}$ is the Hamiltonian constraint and ${\cal H}^i$ the momentum 
constraints and the lapse and shift functions appear only as Lagrange 
multipliers. The functional form of eq.\ (\ref{j5}) is identical 
to the standard Hamiltonian form of the gravitational action in GR 
and the Hamiltonian
\begin{equation}\label{j7}
H = \int d^dx \, \Big[N {\cal H} +N_i {\cal H}^i\Big]
\end{equation}
is proportional to the constraints like in GR.

In the following we will consider only 1+1 dimensions.
In addition we restrict ourselves to so-called projectable HL gravity,
where one assumes that the lapse function $N(x,t)$ depends 
only on $t$. This restriction is compatible with the foliation
preserving diffeomorphisms. In 1+1 dimensions $\kappa$ is a dimensionless
coupling constant and it will play no role. We thus set $\kappa =1$
in the following. The 1+1 dimensional action simplifies dramatically. 
We have $R=0$ since $d=1$ and further $g_{11} = g$. 
It is convenient to use as canonical variable 
$\g = \sqrt{g}$:
\begin{equation}\label{j7a}
I= \int dt \; dx \; 
N \g \left[ (1-\lambda) K^{2} - 2 \Lambda \right]\;, 	
\end{equation}
where
\begin{align}\label{j8}
	K&= \frac{1}{N} \left( \frac{1}{\g} \partial_{0} \g - 
\frac{1}{\g^2} \partial_{1}N_{1} + \frac{N_{1}}{\g^3} \partial_{1} \g \right) \;;
\end{align}
The conjugate momentum of our metric variable $\g$ is 
\beq\label{j9}
\pg= \pd{\mathcal{L}}{(\partial_{0}\g)} = 2(1- \lambda) K ,
~~~~\{\g(x,t),\pg(y,t)\} = \del(x-y), 
\eeq
where the last equation states the Poisson bracket between $\g$ and $\pg$.
Eq.\ \rf{j7} now reads:
\beq\label{j10}
H = \int dx \;[N {\cal H} +N_1 {\cal H}^1], ~~~
\cH = \g \frac{(\pg)^2}{4(1-\lam)} +2\Lam \g,~~~\cH^1= -\frac{\prt_1\pg}{\g}.
\eeq
A great simplification comes from the momentum constraint 
\beq\label{j11}
\cH^1=0 ~~~{\rm i.e.}~~~\pg(x,t)=\pg(t).
\eeq
On this constraint surface $H$ can be expressed as follows: 
\beq\label{j12}
H = N(t)\, \Big( L(t)\, \frac{(\pg(t))^2}{4(1-\lam)} +2\Lam L(t)\Big),~~~~
L(t):= \int dx \,\g(x,t),
\eeq
entirely in terms of variables with no $x$ dependence. 
Note that both $L(t)$ and $\pg(t)$ are invariant
under the foliation preserving diffeomorphisms \rf{j1}. 
$N(t)$ can be regarded as the Lagrange multiplier of the 
Hamiltonian constraint. This fixes $\pg(t)$ to 
be a $t$ independent constant provided  $(\lam-1)\Lam > 0$:
\beq\label{j12a}
(\pg)^2 = 8(\lam-1)\Lam.
\eeq
This equation tells us that the constraint surface is a surface 
of constant extrinsic curvature, provided the rhs is 
positive. It clearly depends on  $\lam$ and $\Lam$ whether or not 
that is the case. If the rhs of eq.\ \rf{j12a} is negative
the only solution to the constraint coming from $N(t)$ is $L(t)=0$.  
In this case, if we {\it want} to start with a  dynamical system 
with a genuine non-trivial classical solution, we can 
add the constraint that the volume of spacetime should be a  constant $A$.
Effectively this means that the cosmological constant is treated 
as a  Lagrange multiplier, which after solving the eom is 
fixed such that the corresponding spacetime volume is $A~$\footnote{The same 
problem is encountered in 2d Liouville gravity,  
and the same solution is used: the Liouville equation can
be written as $R = -\Lam$, where $R$ is the 2d scalar curvature.
But since $\int \sqrt{g} R = 2\pi \chi$, $\chi$ being 
the Euler characteristics of the 2d manifold, 
there is no classical solution for positive $\Lam$ and 
$\chi$ not negative unless one e.g.\ fixes the total spacetime volume.}.
However, using the path integral quantization, as we will in 
the next section, there is no formal problem  associated with 
a quantization around $L(t)=0$.
With this understanding the classical dynamics of $L(t)$ 
is entirely determined by the action
\beq\label{j15}
S = \int dt \,  \left( \frac{\dot{L}^2}{4N(t) L(t)} -\tL N(t)L(t)\right),
~~~~~
\tL= \frac{\Lam}{2(1-\lam)},
\eeq
and the corresponding Hamiltonian is 
\beq\label{j16}
H= N(t) \, \Big( L \,\Pi^2 + \tL L\Big),~~~~\{L(t),\Pi(t)\} =1.
\eeq

\section{Quantization}

In order to make contact to CDT we now perform a formal 
rotation to Euclidean signature and use the path integral 
to quantize the theory. Thanks to the existence of 
a time foliation in HL gravity we can think of this as 
a formal substitution $t \to it$ which changes the sign of the 
first term in \rf{j1}\footnote{Of course the metric functions
appearing in \rf{j1} have in general no such well defined
analytic continuation. The same  is true for fields
$\phi(t,x)$ when one performs the rotation to Euclidean signature:
general fields appearing in the path integral do not   
allow such a rotation. In fact they are with probability one 
not even analytic functions of $t$. However, certain 
amplitudes, calculated in flat Lorentzian spacetime can be rotated
to amplitudes calculated in Euclidean spacetime.}.
Note that in projectable HL gravity proper time
\beq\label{j16a}
t_p(t) := \int_0^t dt' N(t')
\eeq
is a physical observable: it cannot be changed by an 
allowed redefinition \rf{j1} of spacetime. Since $L(t)$ is 
also invariant under time redefinition we can, in HL gravity,
view $L(t_p)$ as an observable developing in ``real'' observable
time $t_p$. In the quantum theory it is thus a perfectly 
legitimate question to ask for the probability amplitude
that the universe evolves from a state of length $L_1$ 
(i.e.\ an eigenstate of the operator $\hL$ with eigenvalue
$L_1$) to a state of length $L_2$ in proper time $T$.
This amplitude $G(L_2,L_1;T)$ is precisely given by the path integral
(we give the Euclidean version)  
\beq\label{j17}
G(L_2,L_1;T) = \int \frac{\cD N(t)}{\mathrm{Diff[0,1]}} 
\int \cD L(t) \; \e^{ -S_E[N(t),L(t)]}.
\eeq
Here $S_E$ is the action \rf{j15} rotated to Euclidean signature
\beq\label{j17a}
S_E = \int dt \,  \left( \frac{\dot{L}^2}{ 4N(t) L(t)} +\tL N(t)L(t)\right),
\eeq
and we integrate over all functions 
\beq\label{j18}
N:~t \to N(t),~~t\in [0,1],~~~\int_0^1 dt \,N(t) =T,
\eeq
\beq\label{j19}
L:~t \to L(t),~~t\in [0,1],~~L(0)=L_1,~~L(1)=L_2.
\eeq
By standard gauge fixing to proper time $N(t)$ drops out of the
action and any reference to $N(t)$ disappears in \rf{j17}. We are thus 
left with an ordinary path integral where the time is proper time,
which, as discussed above, is an observable in HL gravity.  
$G(L_2,L_1;T)$ defines the quantum Hamiltonian $\hH$ by:
\bea\label{j20}
G(L_2,L_1;T) &=& \la L_2| \,\e^{-T \hH}\, |L_1\ra,\\
G(L_2,L_1;\ep) &=& \la L_2| \,(\hat{1} -\ep \hH + O(\ep^{3/2}))\, |L_1\ra.
\label{j21}
\eea
In order to extract $\hH$ from \rf{j20} one uses the basic 
definition of the path integral. The proper time interval 
$[0,T]$ is subdivided in pieces of length $\ep$, 
(assuming $\cD L(t) = \prod dL(t_i)$) 
\beq\label{j22}
\int_0^\infty dL_1G(L_2,L_1;\ep)\psi (L_1) = \psi(L_2) -\ep (\hH \psi)(L_2) 
+O(\ep^{3/2}).
\eeq
The result of this standard calculation is:
\beq\label{j23}
\hH = - \frac{d}{dL} L \frac{d}{dL} + \tL L.
\eeq
This is precisely the CDT Hamiltonian derived in \cite{dgk} and 
\cite{durhuus}. 

A much simpler way to derive the Hamiltonian is to gauge fix \rf{j16}
by choosing proper time. Then we have  classically $H= L \Pi^2 +\tL L$, and 
\beq\label{j24}
\{ L,\Pi\} =1 \to [\hL, \hat{\Pi}] =i~~~{\rm by}~~~~
\hat{\Pi}= -i \frac{d}{dL}.
\eeq    
However, this procedure does not resolve the operator ordering 
ambiguity that arises when  $L \Pi^2$ is promoted to a quantum operator. 
In fact we have three options
\beq\label{j25}
 (L\Pi^2)_0 =  - \frac{d}{dL} L \frac{d}{dL},~~~
(L\Pi^2)_{-1} =  -L \frac{d^2}{dL^2},~~~
(L\Pi^2)_{1} =  - \frac{d^2}{dL^2} L 
\eeq 
Option $0$ leads to an operator which is Hermitian
when we use the flat measure $dL$ on the positive real axis
(and appropriate boundary conditions). Option $-1$ leads to 
a Hermitian operator if we choose the measure $dL/L$, and
similarly option $1$ leads to a Hermitian operator when 
we use $LdL$ (all with appropriate boundary conditions, of course). 
For these different choices of Hamiltonians, $\hH_i$, $i=-1,0,1$, the  
amplitudes $G_i(L_2,L_1;T)$ satisfy different composition rules:
\beq\label{j26}
G_i(L_2,L_1;T_1+T_2)= \int_0^\infty dL \, G(L_2,L;T_2) \,L^i \, 
G(L,L_1;T_1)
\eeq   
We can of course derive the Hamiltonians $\hH_i$, $i=\pm 1$ from 
the continuum path integral in the same way as we derived \rf{j23},
provided we start out with the measures 
\beq\label{j27}
\cD L(t) = \lim_{\ep \to 0} \prod_{k=1}^{T/\ep} L^i(t_k) \;dL(t_k).
\eeq
The geometrical interpretation of these measures for $i=\pm 1$
was already given in the original CDT article \cite{al}. The
measure $L dL$ corresponds to considering spatial universes 
which are closed (i.e.\ a circle) and where no points are marked
on the boundaries. 
These universes come with  symmetry factors $1/L$. 
However, when these universes are glued together, as in \rf{j26}, 
the boundaries become internal.
To then correctly count the different surfaces that arise 
from gluing one needs to compensate for the $1/L$ factor 
by multiplying with $L$.
Similarly, the measure $dL/L$ arises from gluing together cylinder 
amplitudes where one has chosen to mark a point
on the boundary. Finally, as shown in \cite{dgk} the flat 
measure $dL$  describes the situation where we have open boundaries 
of length $L$. In this case there is no symmetry factor associated with 
the gluing. It is remarkable that all operator orderings have 
a simple geometric realization.

In our projectable HL gravity the classical Hamiltonian constraint
$L \Pi^2 + \tL L=0$ on the constraint surface leads to the quantum
statement that the Hartle-Hawking wave function for a closed universe 
satisfies 
\beq\label{j30}
\hH_{\pm 1} W^{(\pm 1)}(L) =0, ~~~\hH = \hat{L\Pi^2}_{\pm 1} +\tL \hL
\eeq 
depending on how we treat the boundary conditions. Eq.\ \rf{j30}
is nothing but the Wheeler-DeWitt equation.   
The Hartle-Hawking  ground state of the universe describes the ``propagation''
in Euclidean spacetime from ``nothing'' to a given space of length $L$. 
This is what in the context of CDT has been called the ``disk'' function
since the topology of spacetime is that of a disk if we start
from ``nothing''. Explicitly
the disk functions are \cite{al} 
\beq\label{j31}
W^{(\pm 1)}(L) = \int_0^{\infty} dT \; G_{\pm 1}(L,L_1=0;T);
\eeq
\beq\label{j32}
W^{(1) }(L)
=\e^{-\sqrt{\tL} L}/L,~~~~W^{(-1)}(L)= \e^{-\sqrt{\tL} L}.
\eeq
Although the disk functions $W^{(\pm 1)} (L)$ 
satisfy $\hH_{\pm 1} W^{(\pm 1)} (L) = 0$ they are not 
normalizable wavefunctions with respect to the measures $dL L^{\pm 1}$,
respectively (nothing tells us that the wavefunction
of the universe has to be normalizable!). 
The energy spectra of $\hH_{\pm 1}$ are positive 
definite (for $\tL >0$) with eigenvalues $E_{\pm1}(n) = 2\sqrt{\tL} (n+1)$,
$n=0,1,\ldots$ (see \cite{topologies} where also the Hartle-Hawking 
wave function  which includes the  summation over spacetime 
topologies can be found).

\section{Discussion}

We have identified the continuum theory corresponding to 
2d CDT. It is projectable 2d HL gravity with $\lam <1$ and $\Lam >0$.
By expressing the HL theory in terms of 
proper time $T$ and the scale factor of the universe $L(T)$,
and then quantizing the theory, we obtained a quantum theory 
identical to the continuum limit of CDT.

2d HL gravity offers us an interesting laboratory for addressing
the standard conundrum of rotating to the Euclidean sector: if we
choose $\lam >1$ we imitate the standard situation we believe is 
realized in our universe for positive $\Lam$: 
it is expanding (or contracting) exponentially in (proper) time $T$.
However, the ``kinetic'' scale factor term which leads to this 
expansion has the wrong sign compared to an ordinary kinetic 
term (precisely as in ordinary GR), 
causing disaster when {\it naively} rotated to the Euclidean 
sector. If we choose $\lam <1$ the situation is healthy 
compared to an ordinary dynamical system, but it does not
correspond to a dynamical system resembling our universe unless
we choose $\Lam <0$. At a classical level there is no
difference since we can just change the sign of $S$ as we 
have done by unifying the situations in \rf{j15}. However,
when coupling to matter and in higher dimensions where 
one might also have transverse physical {\it field} degrees 
of freedom one might not have that option. This needs
to be analyzed. 

It might be interesting to analyze  non-projectable 
2d HL gravity. In this case one has the  ``many'' fingered 
proper time of Wheeler: 
\beq\label{j40}
t_p(t,x) = \int_0^t dt' \, N(t',x),
\eeq
which allows us to consider more general amplitudes.
It is tempting to conjecture that the quantum theory 
corresponding to non-projectable 2d HL is so-called
generalized CDT \cite{GCDT}, which also allows
more general amplitudes. Generalized CDT
was originally formulated in the continuum and  
has a continuum string field theory representation \cite{CDTSFT}
in the same way as ``ordinary'' dynamical triangulation (DT) has 
\cite{SFT}. Later a discretized 
representation of generalized CDT was found and it is universal in the 
same way as DT (and has a matrix model representation in the 
same way as DT \cite{matrix}). More specifically any reasonable class of 
so-called planar maps has as its continuum limit 
Euclidean Liouville quantum gravity. It has recently 
been shown that (continuum) generalized CDT is described as 
the continuum limit of any reasonable class of planar maps
with a finite number of faces \cite{ab}. So we understand
in quite some detail both the regularized theory and its
continuum limit, and what is lacking is a continuum 
classical theory which when quantized {\it is} generalized
CDT. Non-projectable 2d Ho\v rava-Lifshitz gravity is 
a candidate.

Finally, to avoid confusion, 
let us comment on the relation between 2d HL gravity and 
2d GR. Since the Einstein action is topological in 2d we 
have to define what we mean by 2d GR. Here we will think 
about 2d GR as the quantum theory defined by Liouville gravity,
i.e.\ in the Euclidean context the theory which results after 
integrating over all geometry of a fixed topology. It can 
still depend on diffeomorphism invariant boundary data, i.e.\
the length of the boundaries. However, in this theory there is 
no propagating degrees of freedom (the Liouville field is not 
a gravitational field degree of freedom) and the only scale is 
set by the cosmological constant. The same is true for CDT or 
quantized 2d HL gravity. Since we are only considering theories
with quadratic derivative terms (both theories are renormalizable
without adding higher derivative terms) there is no ``flow'' 
of 2d HL to an isotropic point, as is the case in higher 
dimensions. It does not mean that the two theories are 
not related. In fact there exists an explicit map between the 
theories \cite{ackl} and one obtains CDT by integrating out 
so-called baby universes present in DT. Generalized CDT presents 
in some sense an interpolation between CDT and DT and again 
it is tempting to suggest that non-projectable quantum HL can provide 
us with such an interpolation with the parameter $\lambda \to 1$
being the limit where Liouville quantum gravity should be 
recovered.

\vspace{.5cm}      

\noindent {\bf Acknowledgments.} 
LG would like to thank Lu\'{i}s Pires for helpful discussions. 
JA and LG were partly supported by  
the Danish Research Council via the grant ``Quantum gravity and the role of Black holes''. 
JA and YW were partly supported by the 
the EU through the ERC Advanced
Grant 291092, "Exploring the Quantum Universe" (EQU).  
 YS and YW  thank the Niels Bohr Institute for  warm hospitality.
YS is supported by the Grant-in-Aid for Nagoya University 
Global COE Program, ``Quest for Fundamental Principles in the Universe: 
from Particles to the Solar System and the Cosmos".

\end{document}